\documentclass[iop]{emulateapj}

\usepackage{CJK}
\usepackage{txfonts}
\usepackage{graphicx}

\usepackage{graphicx}

\usepackage{soul}

\usepackage{natbib}
\usepackage{natbib}

\begin{document}
\title{Spectroscopic inversions of the C\lowercase{a} \lowercase{{\sc{ii}}} 8542 {\AA} line in a C-class solar flare}
\vskip1.0truecm
\author{D. Kuridze$^{1,5}$, V. Henriques$^{1,2}$, M. Mathioudakis$^{1}$, J. Koza$^3$, T. V. Zaqarashvili$^{4,5,6}$,  J. Ryb\'ak, A. Hanslmeier$^{4}$ and F. P. Keenan${^1}$}  
\affil{$^1$Astrophysics Research Centre, School of Mathematics and Physics, Queen's University Belfast, Belfast BT7~1NN, UK; e-mail: d.kuridze@qub.ac.uk} 
\affil{$^2$Institute of Theoretical Astrophysics, University of Oslo, P.O. Box 1029 Blindern, NO-0315 Oslo, Norway}
\affil{$^3$Astronomical Institute, Slovak Academy of Sciences, 059 60 Tatranska Lomnica, Slovakia}
\affil{$^4$IGAM, Institute of Physics, University of Graz, Universit\"atsplatz 5, 8010 Graz, Austria}
\affil{$^5$Abastumani Astrophysical Observatory at Ilia State University, 3/5 Cholokashvili avenue, 0162 Tbilisi, Georgia}
\affil{$^6$Space Research Institute, Austrian Academy of Sciences, Schmiedlstrasse 6, 8042 Graz, Austria}
\date{received / accepted }

\begin{abstract}

We study the C8.4 class solar flare SOL2016-05-14T11:34 UT
using high-resolution spectral imaging in the Ca {\sc{ii}} 8542 {\AA} line 
obtained  with the CRISP 
imaging spectropolarimeter on the Swedish 1-m Solar Telescope.  
Spectroscopic inversions of the Ca {\sc{ii}} 8542 {\AA} line using the non-LTE code NICOLE are used  
to investigate the evolution of the temperature and velocity structure in the flare chromosphere. 
A comparison of the temperature stratification in flaring and non-flaring areas 
reveals strong footpoint heating during the flare peak in the lower atmosphere. 
The temperature of the flaring footpoints between continuum optical depth at 500~nm,
$\mathrm{log~\tau_{500}~\approx -2.5~and~ -3.5}$
is $\mathrm{\sim5-6.5~kK}$, close to the flare peak, reducing gradually to $\mathrm{\sim5~kK}$. 
The temperature in the middle and upper chromosphere, between $\mathrm{log~\tau_{500} \approx - 3.5~and~- 5.5}$, is estimated to be 
$\mathrm{\sim6.5 - 20~kK}$, decreasing to pre-flare temperatures, $\mathrm{\sim5 - 10~kK}$, after approximately 15 minutes. However, the
temperature stratification of the non-flaring areas is unchanged. 
The inverted velocity fields show that the flaring chromosphere is dominated by
weak downflowing condensations at   
the Ca {\sc{ii}} 8542 {\AA} formation height.

\end{abstract}


\section{Introduction}

It is now widely accepted that the chromosphere is a key to our understanding of solar flares due to the large amount of  radiative losses that originate in it  \citep{2011SSRv..159...19F}. 
Chromospheric radiation  
can provide vital diagnostics for the structure and dynamics of plasma parameters, such as 
temperature, velocity, density, pressure and magnetic field in the flaring atmosphere. 

The method of choice for studies of the one-dimensional vertical stratification of solar and stellar atmospheres has been the use of  
semi-empirical models that attempt to reproduce the observed profiles in LTE or  non-LTE radiative transfer \citep[e.g. see the review by][]{2007ASPC..368..203M}.
Such an approach has been successful for the quiet-Sun (QS) atmosphere, for which static 1D models, developed under an assumption of hydrostatic equilibrium,
reproduce chromospheric spectral lines and continua \citep{1971SoPh...18..347G,1981ApJS...45..635V,1990ApJ...355..700F,1991ApJ...377..712F,1993ApJ...406..319F,2006ApJ...639..441F,2009ApJ...707..482F,2012A&A...540A..86R}.

The first set of static semi-empirical models of a flaring photosphere and chromosphere was developed by \cite{1975SoPh...42..395M}.
These authors modelled the wings of the Ca {\sc{ii}} H\&K lines and concluded that the temperature minimum region in flares  
is both hotter and formed deeper in the atmosphere than the QS models.   
\cite{1980ApJ...242..336M} developed models for a bright and a faint flare,
that reproduce lines and continua of H {\sc{i}}, Si {\sc{i}}, C {\sc{i}}, Ca {\sc{ii}}, and Mg {\sc{ii}}, and
show a substantial temperature enhancement from the photosphere up to the transition region.  
Semi-empirical models of a white-light flare also show that flare related perturbations can affect the wing and continuum formation heights, as well as the continuum emission originating in the photosphere \citep{1990ApJS...74..609M,1990ApJ...360..715M,1993ApJ...414..928M}. 

\cite{1987SoPh..107..311G} and \cite{1993ApJ...416..886G} used H$\alpha$ line profiles to construct semi-empirical models for two flares
which showed evidence for chromospheric condensations, which can  
reproduce the well-observed H$\alpha$ line profile asymmetries in flares. 
Moreover, \cite{1994ApJ...430..891G} found that condensations can increase the back-warming of the atmosphere, leading to the heating of the photosphere and 
enhancement of the continuum emission. The evolution of chromospheric velocity fields during solar
flares was also studied by \cite{2002A&A...387..678F} and \cite{2005A&A...430..679B}. 
\cite{2002A&A...387..678F} constructed 5 semi-empirical models for different flare evolution times, 
which reproduce the profiles of the H$\delta$, Ca {\sc{ii}} K and Si {\sc{i}} 3905\,\AA\ lines. 
Velocity fields were included to reproduce the asymmetric line profiles.
\cite{2008A&A...490..315B}  performed semi-empirical modeling of the solar flaring atmosphere above sunspots using NLTE radiative
transfer techniques. 
They found that the flaring layers (loops) are dominated by chromospheric evaporation leading to a significant increase of gas pressure.
Flare models obtained with forward radiative hydrodynamic codes such as RADYN \citep{1997ApJ...481..500C}
can also reproduce the observed asymmetric line profiles and temperature increase in the lower solar atmosphere \citep{2015ApJ...813..125K,2016ApJ...832..147K}.

A powerful way to construct atmospheric models with a semi-empirical approach, is to fit the observed Stokes profiles using 
inversion algorithms \citep{2016SSRv..tmp...73D}. 
Through such inversions, the ionisation equilibrium, statistical equilibrium and radiative transfer equations are solved numerically to 
synthesize the Stokes profiles under a set of pre-defined initial atmospheric conditions.
Differences between the observed and synthetic profiles are used to modify, at a height usually determined by a response function, an  initial model atmosphere that is used to reproduce the observed spectral signatures. 
 
Several inversions have been performed in the Ca {\sc{ii}} H line using the LTE Stokes Inversion based on Response functions code
\cite[SIR,][]{1992ApJ...398..375R}.    
A first-order NLTE correction was applied to obtain the temperature stratification of the QS and active region chromosphere \citep{2013A&A...553A..73B,2015ApJ...798..100B}.
However, the development of NLTE radiative transfer codes, such as HAZEL
\citep{2008ApJ...683..542A}, HELIX$^+$ \citep{2009ASPC..415..327L} and NICOLE \citep{2015A&A...577A...7S} 
combined with the increased computational power available, also make NLTE inversions possible (see the review by \cite{2016SSRv..tmp...73D}).  

The Ca {\sc{ii}} infrared (IR) triplet line at 8542\,\AA\ is well suited for the development of chromospheric models due 
to its sensitivity to physical parameters, including magnetic field, in the solar photosphere and chromosphere  
\citep{2008A&A...480..515C,2007ApJ...663.1386P,2016MNRAS.459.3363Q}.
Furthermore, the optimization of NICOLE for Ca {\sc{ii}} 8542 {\AA} 
allows the use of this feature for the computation of semi-empirical atmospheric models.  
For more details about the spectropolarimetric capabilities of Ca {\sc{ii}} 8542 {\AA} the reader is referred to \cite{2016MNRAS.459.3363Q}. 
 
Inversions with NICOLE have been successfully performed for 
spectro-polarimetric Ca {\sc{ii}} 8542 {\AA} observations of umbral flashes in sunspots \citep{2013A&A...556A.115D}, and granular-sized magnetic elements (magnetic bubbles) 
in an active region \citep{2015ApJ...810..145D}.
In this paper we use NICOLE to invert high-resolution Ca {\sc{ii}} 8542 {\AA} spectral imaging observations 
to construct models for the lower atmosphere of a solar flare. Multiple inversions were performed for the observed region covering flare ribbons and for non-flaring areas at different times during the event. From the constructed models we investigate the structure, 
spatial and temporal evolutions of basic physical parameters in the upper photosphere and chromosphere during the flare.  


\section{Observations and data reduction}
\label{sect:setup}

\begin{figure*}[t]
\begin{center}
\includegraphics[width=18.1cm]{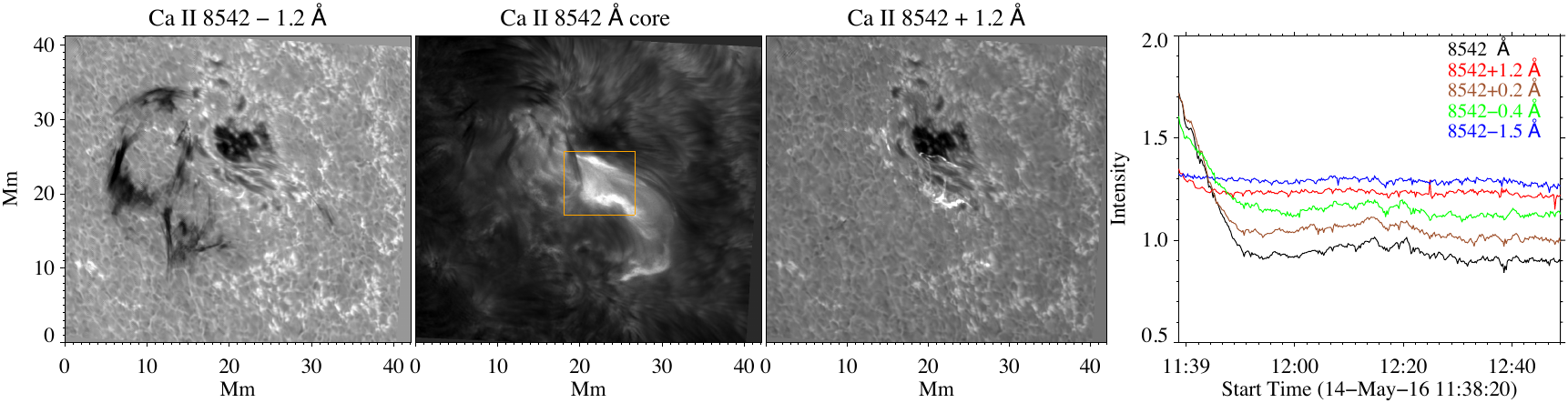}
\end{center}
\caption{The Ca {\sc{ii}} 8542 ${\AA}$ line wing and core images obtained with the CRISP instrument on the SST at 11:38:20 UT on 
2016 May 14. The orange box indicates the flaring region analysed in this paper. The temporal evolution 
of the region averaged over the area marked with the yellow box is presented in the far right panel.}
\label{fig1}
\end{figure*}

The observations were undertaken between 11:38 and 12:49 UT on 2016 May 14 
close to the West limb ($877''$, $-66''$), near the equator 
with the CRisp Imaging SpectroPolarimeter \citep[CRISP;][]{2006A&A...447.1111S,2008ApJ...689L..69S} 
instrument, mounted on the Swedish 1-m Solar Telescope \citep[SST;][]{2003SPIE.4853..341S}
on La Palma. Adaptive optics were used throughout the observations, consisting of a tip-tilt mirror 
and a 85-electrode deformable mirror setup that is an upgrade of the system described in \cite{2003SPIE.4853..370S}. 
The observations comprised of spectral imaging in the H$\alpha$~6563~{\AA} and Ca {\sc{ii}} 8542 ${\AA}$ lines. 
All data were reconstructed with Multi-Object Multi-Frame Blind Deconvolution \citep[MOMFBD;][]{2005SoPh..228..191V}. 
The CRISP instrument includes three different cameras. One camera acquires wideband (WB) images directly from the prefilter and two narrowband (NB) cameras placed after a 50/50 polarizing beam splitter
acquire narrowband (transmitted and reflected) horizontal and vertical polarization
components, each combination of camera, wavelength and component being an object for reconstruction 
\cite[for setup details see e.g.][]{2015A&A...573A..40D}. 
The WB channel provides frames for every time-step, synchronously with the  other cameras, and can be used as an alignment reference.
The images, reconstructed from the narrowband wavelengths, are aligned at the smallest scales by using the method described by \cite{2013PhDT.........2H}. 
This employs cross-correlation between auxiliary wideband channels, obtained from an extended MOMFBD scheme, to account for different residual small-scale seeing distortions. 
We applied the CRISP data reduction pipeline as described in \cite{2015A&A...573A..40D} which includes small-scale seeing compensation as in \cite{2012A&A...548A.114H}.
Our spatial sampling was 0$''$.057 pixel$^{-1}$ and the spatial resolution was close to the diffraction limit of the telescope for many images in the time-series over the $41\times41$\,Mm$^2$ field-of-view (FOV).
For the H$\alpha$ line scan we observed in 15 positions symmetrically sampled from the line core in 0.2 ${\AA}$ steps. The
Ca {\sc{ii}} 8542 ${\AA}$ scan consisted of 25 line positions ranging from $\mathrm{-1.2~{\AA}~to~+1.2~{\AA}}$ from the line core 
with 0.1 ${\AA}$ steps, plus 1 position at $\mathrm{-1.5~{\AA}}$.
A full spectral scan for both lines had a total acquisition time of 12 s, which is the temporal cadence of the time-series.
However, we note that the present paper includes only the analysis of the Ca {\sc{ii}} 8542 {\AA} data which had a duration of 7\,seconds per scan. 
The transmission FWHM for Ca {\sc{ii}} 8542 ${\AA}$ is 107.3 m{\AA} with a pre-filter FWHM of 9.3 {\AA} \citep{2015A&A...573A..40D}.
A two-ribbon C8.4 flare was observed in active region NOAA 12543 during our observations.
Throughout the analysis we made use of CRISPEX \citep{2012ApJ...750...22V}, a versatile widget-based tool for effective viewing and exploration of multi-dimensional imaging spectroscopy data.

Figure~\ref{fig1} shows a sample of the flare images in the Ca {\sc{ii}} 8542 $\AA$ line core and wing positions.
Our observations commenced 4 minutes after the main flare peak ($\sim$11:34 UT). Light curves generated from the region marked with the orange box in Figure~1 
show the post-peak evolution of the emission in the line core and wings 
(right panel of Figure~\ref{fig1}). Although we missed the rise phase and flare peak, at 11:38 UT two bright ribbons associated 
with post-flare emission are clearly detected with good spatial and temporal coverage.

\section{Inversions}

\begin{figure*}[t]
\begin{center}
\includegraphics[width=17.7 cm]{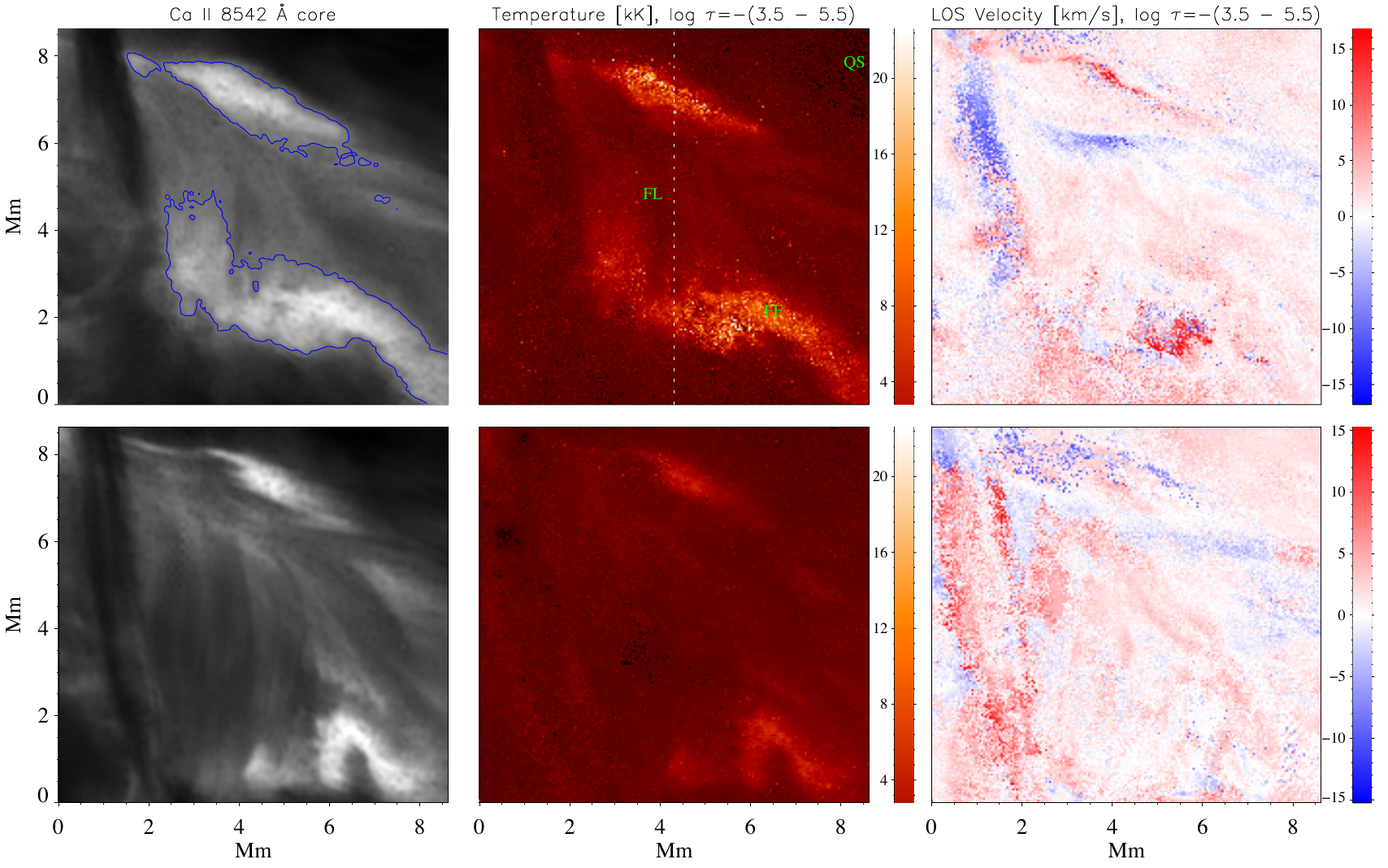}
\caption{The top row shows the images at 11:38 UT (4 minutes after the flare peak) and the bottom row 
images at 11:45 UT. SST images of the flaring region in the Ca {\sc{ii}} 8542 ${\AA}$ line core are shown in the left panels. 
NICOLE outputs showing the temperature and the LOS velocity maps in the interval $\mathrm{log~\tau \sim-3.5~and-5.5}$ are provided
in the middle and right panels, respectively. Blue contours show the area analysed in this paper which has intensity levels greater than 30{\%} of the intensity maximum. 'FL', 'FF', 'QS' 
mark the selected pixels at the flare loop, flare footpoint, and quiet Sun, respectively
discussed in the text in more detail.
The red and blue colors in the dopplerograms represent positive Doppler velocities (downflows) and negative Doppler velocities (upflows), respectively.
The white dotted line indicates the location where vertical cut of the atmosphere is made for detailed analyses.} 
\label{fig2}
\end{center}
\end{figure*}

\begin{figure*}[t]
\begin{center}
\includegraphics[width=17.2 cm]{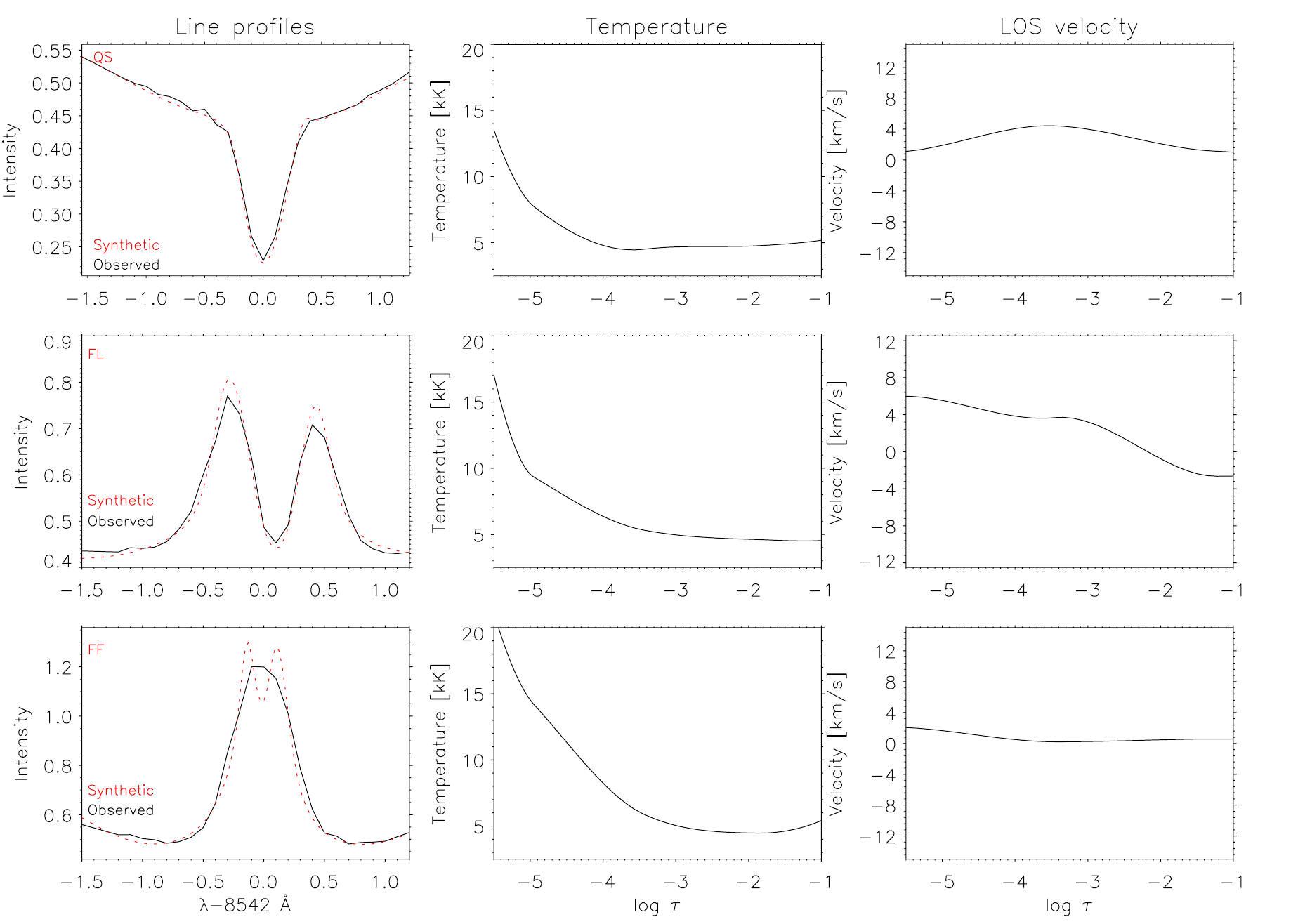}
\caption{Observed (black) and best-fit synthetic (red dotted) Ca\,{\sc ii} 8542\,\AA\ line profiles together with temperature and velocity stratifications for the three selected pixels indicated as FL, FF  
QS in Figure~\ref{fig2}.}
\label{fig3}
\end{center}
\end{figure*}

\begin{table}[b]
\caption{Number of nodes and input atmosphere models used during each cycle (Cy) of the inversion.}
\begin{center}
\begin{tabular}{c   c   c   c  c}
\hline
Physical parameter ~&~~Cy 1~~&~~Cy 2~~&Cy 3& \\  
\hline    
Temperature    &        4 nodes        &         7  nodes  &         7  nodes     \\ 
LOS Velocity    &         1 node             &        5    nodes     &         5  nodes \\ 
Microturbulence    &         1 node             &        1    node     &         1  node \\ 
Macroturbulence    &         none             &          none     &          none\\                 
Input atmosphere    &      FAL-C       &            model from Cy 1   &model from Cy 2        \\
\hline
\label{tabl1}
\end{tabular}
\end{center}
\end{table}

We used the NICOLE inversion algorithm \citep{2015A&A...577A...7S} which
has been parallelized to solve multi-level, NLTE radiative transfer problems \citep{1997ApJ...490..383S}. 
This code iteratively perturbs physical parameters such as  temperature, line-of-sight (LOS) 
velocity, magnetic field and microturbulence 
of an initial guess model atmosphere to find the best match with the observations \citep{2000ApJ...530..977S}. 
The output stratification of the electron and gas pressures, as well as the densities, are computed from the equation-of-state using 
the temperature stratification and the upper boundary condition for the electron pressure under the assumption of hydrostatic equilibrium. 

NICOLE includes a five bound level plus continuum Ca {\sc{ii}} model atom \citep{2009ApJ...694L.128L}
with complete frequency redistribution, which is applicable for lines such as Ca {\sc{ii}} 8542 {\AA}  \citep{1989A&A...213..360U,2016MNRAS.459.3363Q,2011A&A...528A...1W}. 
The synthetic spectra were calculated for a wavelength grid of 113 datapoints in 0.025 {\AA} steps, 
4 times denser than the CRISP dataset. 
Stratification of the atmospheric parameters obtained by the inversions are given as a function of the logarithm of the optical depth-scale at 500~nm (hereafter $\mathrm{log~\tau}$). 
The response function of the Ca {\sc{ii}} 8542 {\AA} line is expanded to
$\mathrm{log~\tau \sim0~and-5.5}$  
\citep[see][for details]{2016MNRAS.459.3363Q}. 
As our Ca~{\sc{ii}}~8542 ${\AA}$ line scan is $\mathrm{-1.5~{\AA}~to~+1.2~{\AA}}$ from line core, 
the analysis of the response functions show that it  provides diagnostics in the layers between $\mathrm{log~\tau\sim-1~and-5.5}$ \citep[see Figure~6 in][]{2016MNRAS.459.3363Q}.

\begin{figure*}[t]
\begin{center}
\includegraphics[width=14.42 cm]{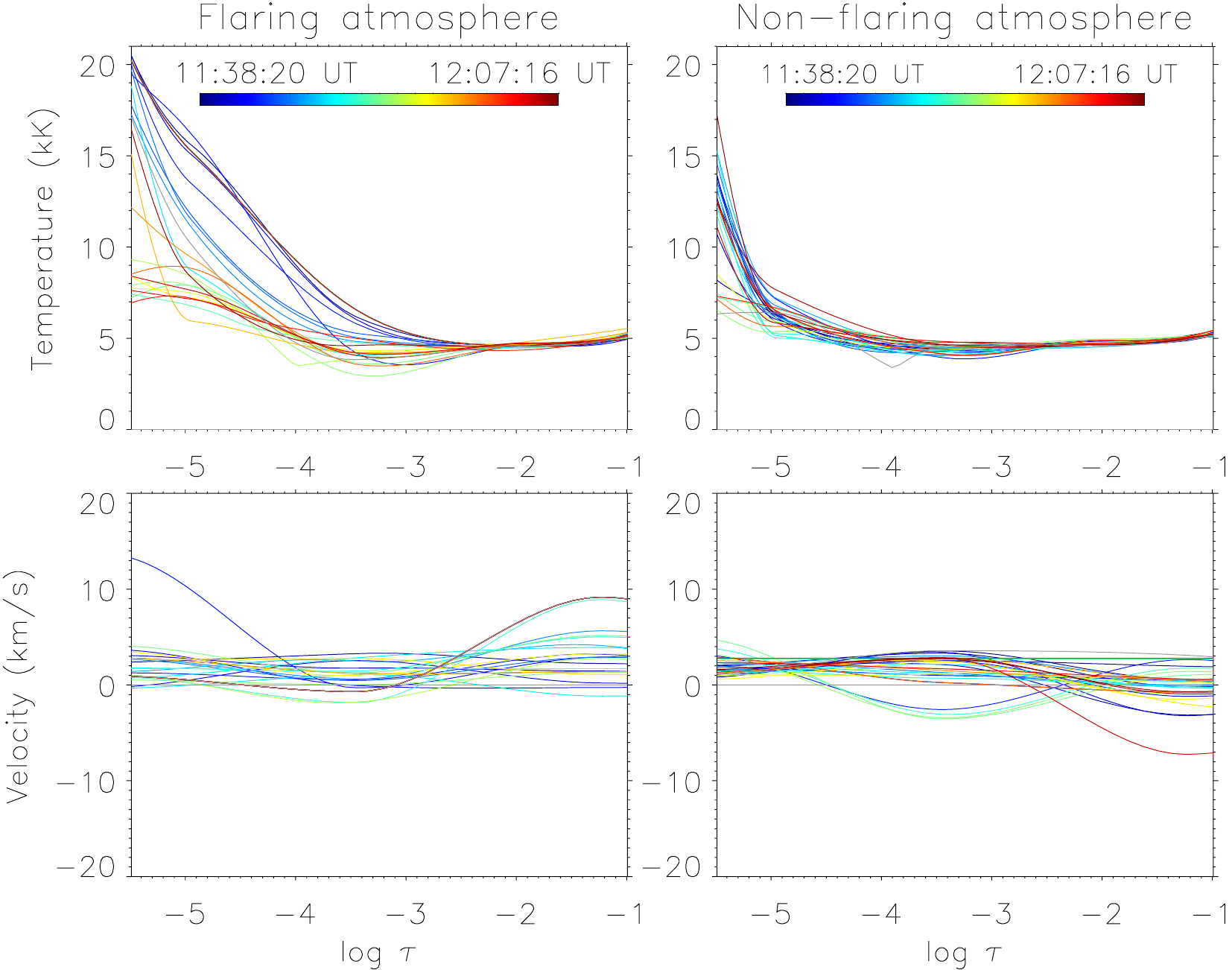}
\caption{The temporal evolution of the temperature and velocity stratifications for the flaring and non-flaring 
pixels marked with QS and FF in Figure~\ref{fig2}, respectively.} 
\label{fig4}
\end{center}
\end{figure*}

To improve convergence, the inversions were performed in three cycles, 
expanding on the suggestion of
\cite{1992ApJ...398..375R} and as implemented 
in \cite{2012A&A...543A..34D}. 
In the first cycle and for the first scan, the inversions were performed with 4 nodes in temperature and 1 in LOS velocity,
with the initial guess model being the FAL-C atmosphere \citep{1993ApJ...406..319F}.
The convergence of the models obtained from the first cycle was improved 
by applying horizontal interpolation to recompute the inverted parameters of the pixels with low quality of fits.
A second cycle was then performed with an increased number of nodes (7 for temperature and 4 for LOS velocity), and the interpolated initial guess model constructed from the first cycle. The synthetic profiles obtained from the second cycle 
tend to have deeper absorptions than the observations, even for the quiet Sun. 
The comparison between
the FTS disk center solar atlas profile (Neckel 1999),
convolved with a model the CRISP transmission
profile with FWHM=107.3 m{\AA} provided by de la Cruz Rodr'guez (2015b),
and the average SST/CRISP flat-field disk-center profile
does not show an adequate coincidence of resulting profiles.
This indicated a broader or asymmetric instrumental profiles than the respective model used in NICOLE (the synthetic profiles are convolved with the instrumental profile before 
comparison with the observations and before the final output). 
Therefore, in the third cycle we convolved 
the instrumental profile with a Gaussian of FWHM=141 m{\AA} and performed an inversion with the same  
number of nodes and an initial guess model constructed from the second cycle.
This led to a better match between the observed and synthetic spectra.
Table 1 summarizes the number of nodes and initial temperature models used in the three cycles.
We also convolved the atlas profile with an
asymmetric Lorentzian transmission profile model
for the CRISP transmission profile
with FWHM = 0.080 {\AA}. This also provides required
similarity between the atlas
and the observed flat-field profiles suggesting that indeed the transmission profile could be asymmetric. 
We run the inversions with the asymmetric Lorentzian profile and the resulted atmospheres were the same
as obtained with the CRISP instrumental profile convolved with a symmetric Gaussian with FWHM = 141 m{\AA}.

%
%

As our observations do not include full Stokes imaging spectroscopic data, 
we run NICOLE only for Stokes I profiles. Therefore, in both cycles, the magnetic field vector and the weights for other Stokes parameters were set to zero. We inverted a 200 $\times$ 200 pixel$^2$ area (8.2$\times$8.2 Mm$^2$)
covering the flare ribbons as well as some non-flaring regions (see left panels of Figure~\ref{fig1}). Although our time-series comprises of 346 spectral scans, we only choose 
the best scans in terms of spatial resolution. 
This limits the total number of scans to 35, with a cadence of approximately 1 minute in the time interval 11:38 to 12:09 UT.


\section{Analysis and results}

\begin{figure*}[t]
\begin{center}
\includegraphics[width=17.9 cm]{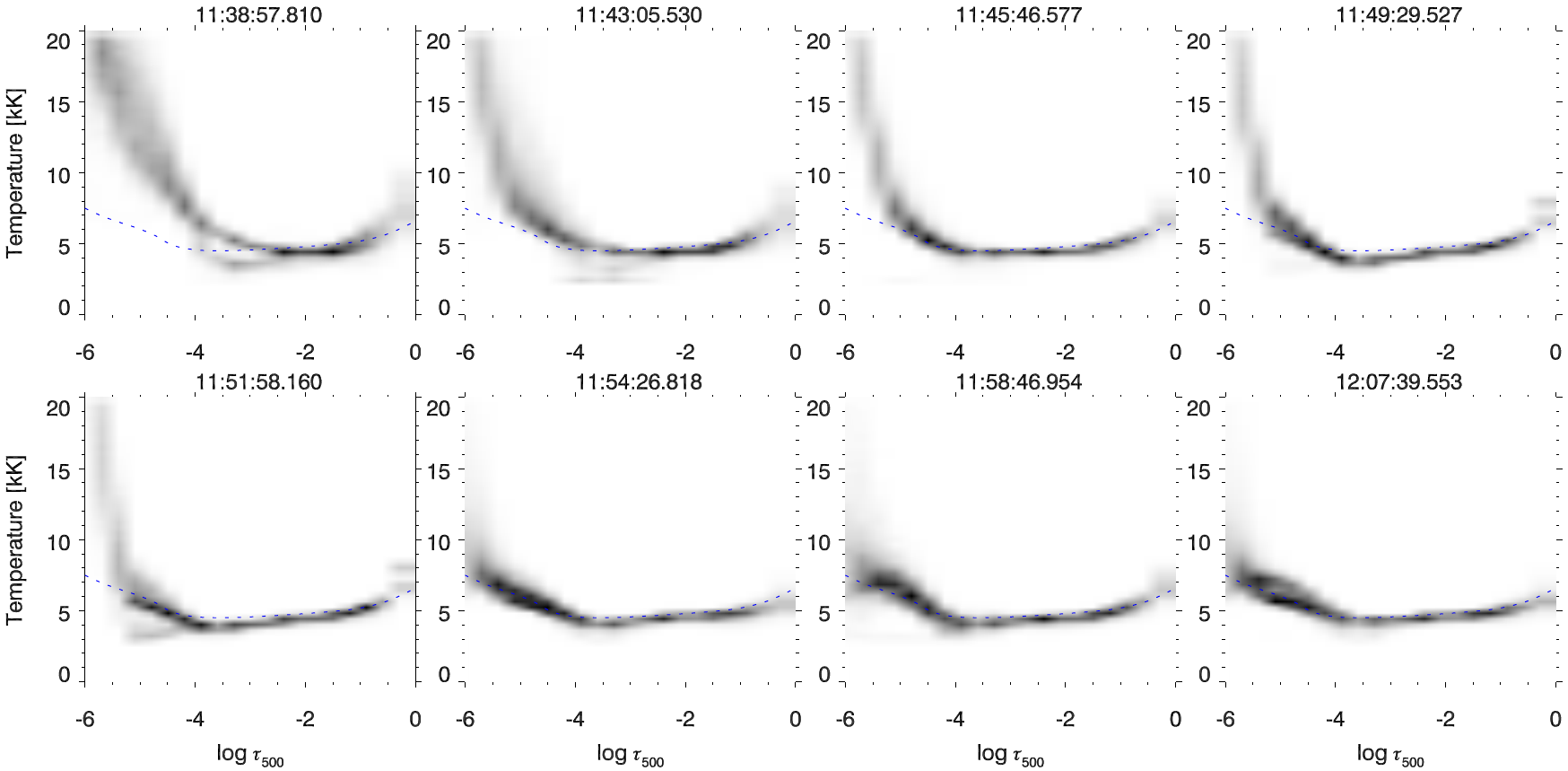}
\caption{Density maps showing the evolution of the temperature stratification during the flare. Each map is produced  
by superimposing the stratification of around 2000 individual models, selected for quality of fit at each time. 
The selected region is marked with blue contours in Figure~2.
The blue dashed line depicts the FALC model, used as the initial atmosphere for the inversions.
} 
\label{fig5}
\end{center}
\end{figure*}

Figure~\ref{fig2} presents a Ca {\sc{ii}} 8542 ${\AA}$ line core image of the region selected for inversions (marked with the orange box in Figure~\ref{fig1}). The NICOLE output showing the temperature and the LOS velocity maps are also presented. 
Temperature and velocity maps are integrated and averaged between $\mathrm{log~\tau\approx-3.5~and-5.5}$, corresponding to  middle and upper chromospheric layers, respectively. The top row shows the images close to the flare peak ($\sim$11:38 UT) 
and the bottom images 7~minutes later. A temperature map of the inverted region close to flare peak 
shows temperature enhancements for the two flare ribbons.  The LOS velocity map of the same region in the top-right panel of 
Figure 2 reveals downflows at the flare ribbons, while the 
temperature map in the bottom-middle panel 
indicates that the size of the flare ribbons and the heated areas have decreased dramatically 7 minutes later.

Figure~\ref{fig3} shows 
the line profiles and temperature/velocity stratifications of 3 pixels indicated as FL, FF, QS in Figure~\ref{fig2}, with the
pixels in different parts of the region under investigation. 
The best-fit synthetic profiles obtained from the inversion at 11:38:20 UT are also shown. There are basically 3 different shapes of line profile over the selected region during the flare: absorption, emission and those with central reversal.  
In the non-flaring areas (QS), Ca {\sc{ii}} 8542 ${\AA}$ shows the well-known absorption line profile. The line profiles of the flaring loops (FL) connecting the flare footpoints (ribbons) have central reversals, whereas 
profiles of the flare footpoints (FF) are in full emission without a central reversal.  
Figure~\ref{fig3} shows that the observed spectra are generally reproduced by the synthetic best-fit spectra. 
The QS temperature is close to the FAL-C temperature profile, whereas the FL has a higher chromospheric temperature at $\mathrm{log~\tau\sim-2.5~and-5.5}$ (middle panels of Figure~\ref{fig3}), with the  
flare footpoints showing the largest temperatures.  

In Figure 4 we present the temporal evolution of the temperature and velocity stratifications for the flaring and non-flaring 
pixels. The temperature of the footpoints is enhanced across the  
chromosphere between optical depths of $\mathrm{log~\tau~\sim -2.5~and~ -5.5}$.
As time progresses, the temperature in the lower chromosphere between $\mathrm{log~\tau~\sim -2.5~and~ -3.5}$ decreases gradually from $\mathrm{T \sim5-6.5~kK}$ to $\mathrm{T \sim5~kK}$. 
In the middle and upper chromosphere, $\mathrm{between~log~\tau \sim -3.5~and~-5.5}$, the temperature decrease from $\mathrm{T \sim6.5 - 20~kK}$ to $\mathrm{T \sim5 - 10~kK}$ during about 15 
minutes (top left panel of Figure~\ref{fig4}). The velocity field for the flaring pixel is dominated by weak upflows which 
decrease gradually with time, while the 
temperature and velocity of the non-flaring areas are lower and unchanged 
in the chromosphere at $\mathrm{log~\tau~\sim -1~and~ -5.5}$ (right panels of Figure~\ref{fig4}). 

\begin{figure*}[t]
\begin{center}
\includegraphics[width=17.4 cm]{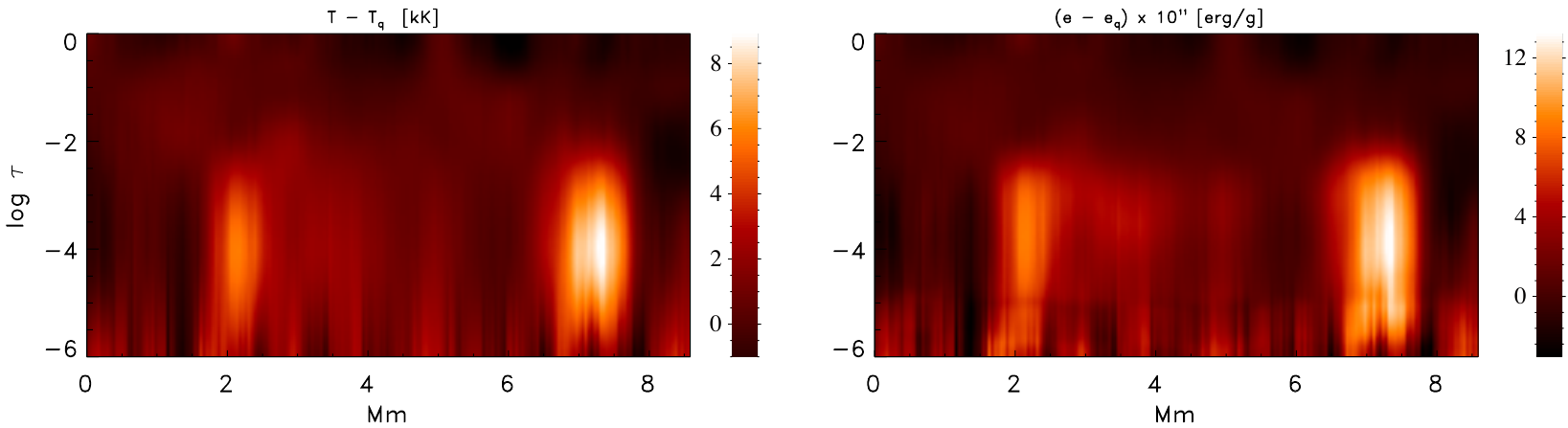}
\includegraphics[width=7.2 cm]{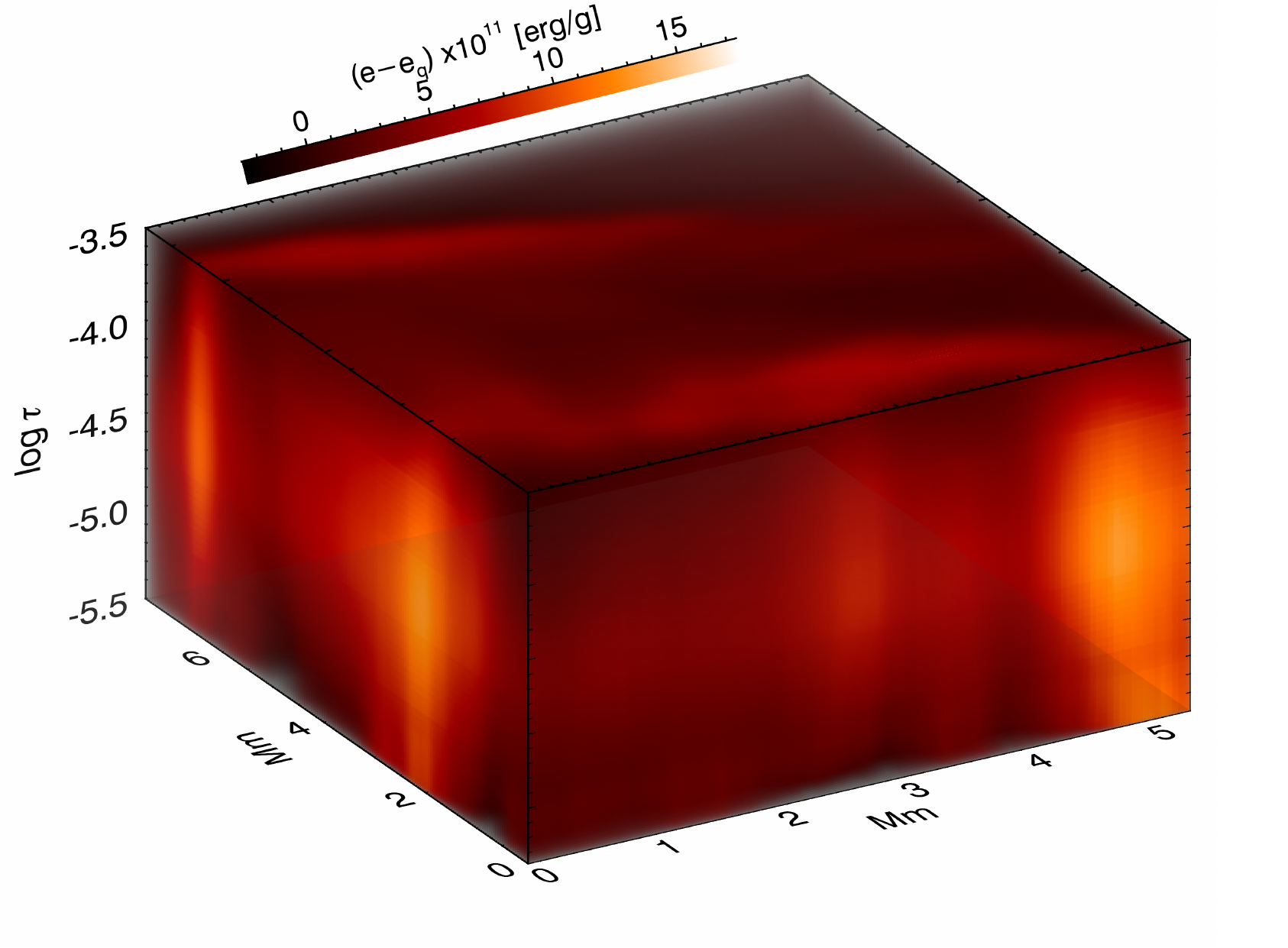}
\includegraphics[width=10.5 cm]{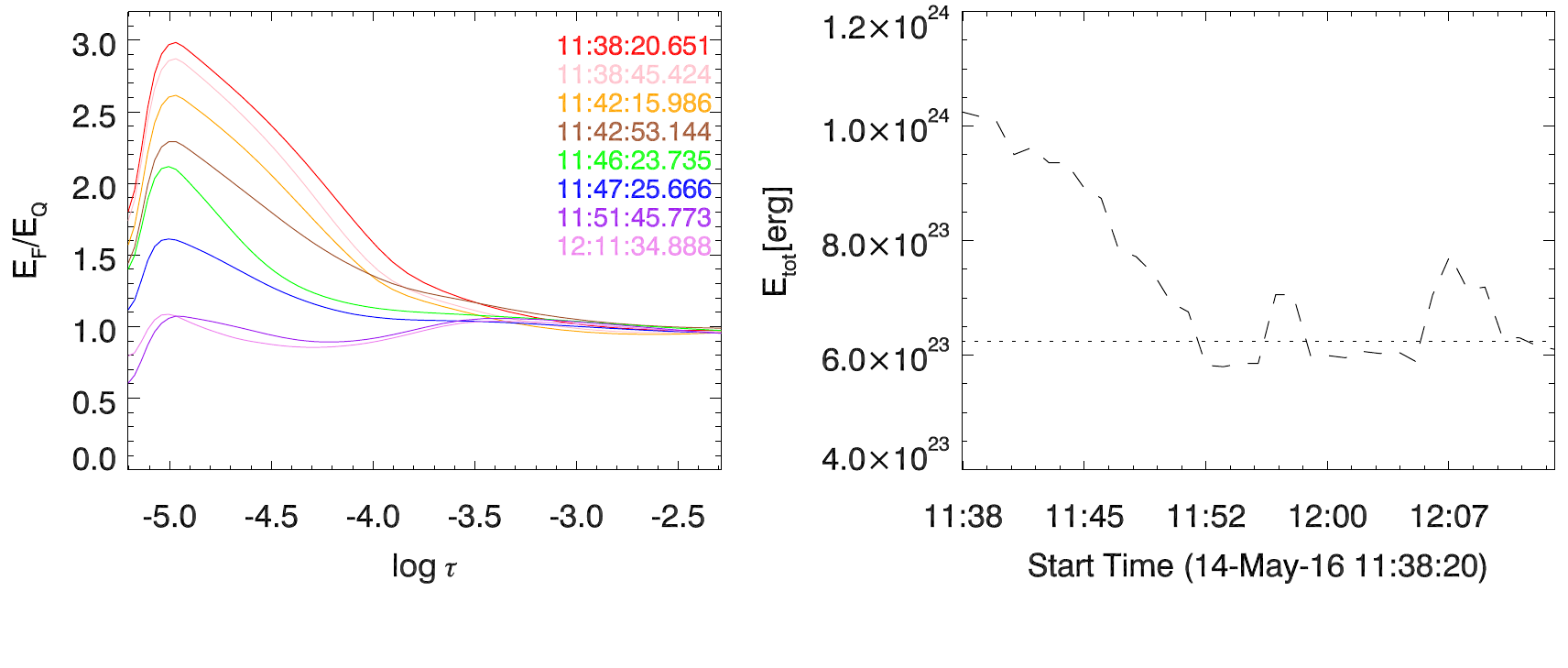}
\caption{Top left: A vertical cut of the temperature stratification with  
the quiet, non-flaring temperature subtracted. The maps are produced along the locations indicated by the white dotted line in Figure~2. Top right: 
The vertical cut of the excess energy density per unit mass, which is the internal energy $e(x,y,\tau)$ computed for each pixel, minus an average internal energy $e_q(x,y,\tau)$ estimated from quiet, 
non-flaring areas. Bottom left: 3D rendering of the excess energy density produced for an inner part (only $S\approx5.3\times6.7$~Mm$^2$) of the inverted datacube. 
The maps have been smoothed over 10 pixels ($\sim$400 km) along the $x$-axis.
Bottom middle: evolution of the ratio of the area-integrated internal energy for the inverted region shown on the bottom left panel over an equally-sized area of QS.
Bottom right: evolution of the total energy (dashed line) 
integrated over
an inverted region presented on the bottom left panel of Figure~6. The total energy of the same-size QS volume is overplotted as dotted line.} 
\label{fig6}
\end{center}
\end{figure*}

In Figure 5 we show the evolution of the temperature 
stratifications using density maps which have been produced  with the superimposition of individual pixels from the flaring region marked with blue contours in Figure 2. 
It must be noted that the stratification of the pixels with low quality of fit, such as non-physically low or high temperature plateaus or dips and peaks, 
were ignored and are not included in the density plots. 
Our density maps confirm that the most intensively heated layers are in the middle and upper chromosphere at  optical
depths of 
$\mathrm{log~\tau~\sim -3.5~and~ -5.5}$, reaching temperature between $\sim$6.5 - 20 kK. 
The temperature stratification of the layers below $\mathrm{log~\tau~\sim -2.5}$ remains unchanged during the flare and is consistent with the QS FALC model used as the initial atmosphere for the inversions (Figure~\ref{fig5}).
In the top-left panel of Figure~\ref{fig6} we show a vertical cut of the net flare temperature enhancement close to the flare peak ($\sim$11:38 UT),
where an average temperature stratification obtained from quiet, non-flaring areas has been subtracted. 
The two bright regions at $\sim$2 and 7.5 Mm show the temperature enhancements of the flare ribbons.  

As noted in Section 3, NICOLE calculates the basic thermodynamical parameters that define the internal energy of the system, such as the density and electron pressure,
using the equation-of-state, temperature stratification and upper boundary condition under the assumption of hydrostatic equilibrium.
To estimate the energy content of the flaring chromosphere, we compute the internal energy {\em e} using the 
relationship:
\begin{equation}
e=\frac{1}{\gamma-1}\frac{p}{\rho},
\end{equation}
where $p$ and $\rho$ are the gas pressure and density, respectively, and $\gamma$=5/3 is 
the ratio of specific heats \citep{2004psci.book.....A,2013A&A...553A..73B}. 
The top right panel of Figure~6 shows a vertical cut of the excess energy density per unit mass close to flare maximum at 11:38 UT,
which is the internal energy, $e(x,y,\tau)$, computed for each pixel minus an average internal energy, $e_q(x,y,\tau)$ estimated from quiet, non-flaring areas. 
A three-dimensional (3D) rendering of this excess energy density for an inner part of the inverted datacube is also shown in the bottom left panel of Figure~6.
It illustrates that the energy density enhancement coincides with the locations of flare ribbons where the temperature enhancements are detected. 
The ratio of total internal energy of this part of the inverted surface
($S\approx5.3\times6.7$~Mm$^2$) over
the internal energy for an equally-sized surface of the QS as a function
of optical depth is presented in the bottom middle panel of Figure~6. 
This shows that the energy of the flaring region is increased 
in the range of optical depths 
log $\mathrm{\tau\sim-2.5-5.5}$ compared to the internal energy of the QS.  
The ratio has a peak at $\mathrm{log~\tau\approx-5}$ where the internal energy of the inverted flare region is increased 
by a factor of cose to 3 over the QS. As time progresses, the ratio decreases gradually to $\sim$1 after about 15 minutes,
meaning that the atmosphere (energetically speaking) has reached QS levels. 
The total energy of the chromospheric volume shown on the bottom left panel of Figure~6 (integrated over the range of optical depth log $\mathrm{\tau\approx-3-5.5}$) 
is estimated to be  $\sim$10$^{24}$ erg close to the flare peak, decreasing linearly to a QS energy level of $\sim\mathrm{6\times10^{23}}$ erg after approximately 15 minutes (bottom right panel of Figure~6).
We note that due to the high viewing angle of the observed region the vertical extend of the model atmospheres shown and described in Figure 6 
does not correspond to the true geometrical height of the lower solar atmosphere. NICOLE derives the 
geometrical height
from optical depth height scale and the temperature plus density/pressure stratifications and
the same optical depth $\tau$ does not refer to the layers at the same geometrical height in the different models.
Therefore, temperature/energy excess derived through the models does not show accurately the energy distribution as a function of geometrical height, but it does so as a function of optical depth.


\section{Discussion and conclusions}

We have presented spectroscopic observations of the Ca {\sc{ii}} 8542\,\AA\ line in a C8.4-class flare. 
The line profiles of the flare ribbons are in total emission without a central reversal, 
whereas those of flaring loops, which connect the flare footpoints, show central reversals (Figure~3). 
In the quiet Sun the line is in absorption (Figure~3).
The analysis of synthetic chromospheric spectral line profiles produced with radiative hydrodynamic models have shown that the changes in temperature, radiation field and population density of the energy states 
make the line profile revert from absorption into emission with or without a central reversal  \citep{2015ApJ...813..125K,2016ApJ...832..147K}.
In a recent investigation, \cite{2016ApJ...832..147K} studied the flare profiles of the Na {\sc{i}} D1 line which also forms in the optically thick chromosphere. 
The heating of the lower solar atmosphere by the non-thermal electron beam makes the Na {\sc{i}} D1 line profiles go into full emission.  However, when the beam heating stops, the profiles develop a central reversal at  the line cores.
Unfortunately, as there is no energy flux from electrons in NICOLE, it cannot be used to investigate how 
the emission and centrally reversed profiles of Ca {\sc{ii}} 8542 {\AA} line are formed. 
However, the analysis of the synthetic line profiles from 
the RADYN simulations shows that for a strong 
electron beam heating (model F11) the Ca {\sc{ii}} 8542\,\AA\ line profiles are in full emission, but develop a central reversal 
after beam heating at the relaxation phase of the simulation \citep{2015ApJ...813..125K}. 
For a weak flare run (model F9), the synthetized Ca {\sc{ii}} 8542\,\AA\ profile exhibits a shallow reversal during the beam heating phase that deepens during the relaxation phase. 

The comparison of synthetic line profiles and their temperature stratifications from different models obtained from the inversions 
presented here indicates that the 
depth of the central reversal depends on the
temperature at the core formation height ($\mathrm{\sim-5.5<\log\tau<-3.5}$), i.e. a higher temperature  
produces a shallower central reversal (Figure~3).
This could explain why flare ribbons, which are believed to be the primary site of non-thermal energy deposition and intense heating, 
have profiles in full emission, 
whereas flaring loops, which are secondary products of the flare and hence less heated areas, have central reversals.

We constructed semi-empirical models of the flaring atmosphere using the spectral inversion NLTE code NICOLE to investigate its structure and evolution.  Models were generated at 35 different times during the flare, starting at 4 minutes after the peak.
The construction of the models is based on a comparison between observed and synthetic spectra. 
Close to the line core, where the self-reversal is formed, we find a small discrepancy between the synthetic and observed spectra 
of the flare ribbons (Figure 3). The synthetic profiles have a small absorption in the core, in contrast to the observed profiles which are in full emission (bottom left panel of Figure~3). This influences the narrow layer of core formation height  
and hence does not have strong effect on the overall output model.
A possible reason of the observed discrepancy could be the lower ionization degree of Ca {\sc{ii}}
in the flaring atmosphere compared to the ionization used in NICOLE \citep{1974SoPh...35...11W}. 
Furthermore, the models constructed close to the flare peak are characterized by high quality of fit and provide a better
match with the observed spectra. This is because, as time from flare maximum progresses, 
the well-defined shapes of emission and centrally-reversed line profiles 
become flatter and feature more irregular shapes, which causes NICOLE to encounter difficulties finding atmospheres that reliably fit such profiles.

Our analysis of the constructed model shows
that the most intensively heated layers in the flaring lower atmosphere are the middle and upper chromosphere at 
optical depths of $\mathrm{log~\tau~\sim -3.5~and~ -5.5}$, respectively,  with temperatures between $\sim$6.5 - 20 kK. 
The temperatures of these layers are decreasing down to typical QS values ($\sim$5 - 10 kK) after about 
15 minutes (Figure~\ref{fig5}). 
In the photosphere, below $\mathrm{\log~\tau\approx-2.5}$, there is no significant difference between quiescence and flaring temperature stratifications (Figure~5).  
This agrees with some of the previous results (e.g. \cite{2002A&A...387..678F}), which show that during the flare the atmosphere is unchanged below $\sim$600 km. 
However, it must be noted that as the observations presented in this work started 4 minutes after the flare peak, it is possible that the  temperature in the deeper layers of the atmosphere was affected during the impulsive phase.  

The velocity field indicates that the observed field-of-view (FOV) is dominated by weak downflows  
at optical depths of $\mathrm{log~\tau~\sim -1~and~ -5.5}$ associated with the post-flare chromospheric condensations (Figures 2 and 3).
Centrally-reversed Ca {\sc{ii}} 8542 $\AA$ profiles show excess emission in the blue wing (blue asymmetry) with a red-shifted line center (middle left panel of Figure~3).
Similar to blue asymmetries observed in H$\alpha$,  Na {\sc{i}} D1 and Mg {\sc{ii}} flaring line profiles \citep{1999ApJ...521..906A,2015ApJ...813..125K,2016ApJ...832..147K,2016ApJ...827..101K}, 
the asymmetry found in Ca {\sc{ii}} 8542\,\AA\
seems to be related to the velocity gradients associated with the chromospheric condensations.  
The core of the  Ca {\sc{ii}} 8542 $\AA$ profile presented in the left middle panel of Figure~3 
is red-shifted owing to the downflows at the height of core formation (middle right panel of Figure~3). 
Velocity decreases downward toward the wing formation regions, producing a
positive velocity gradient with respect to the height inward. This gradient can modify the optical depth of the atmosphere in such a way that higher-lying (core) 
atoms absorb photons with longer wavelengths (red wing photons), 
and the blue asymmetry in the centrally-reversed peak is formed.
We note that semi-empirical models of the flaring atmosphere above sunspots \cite{2008A&A...490..315B}  
indicate that at the early phase of the flare the flaring layers (loops) are dominated by chromospheric upwlows (evaporations) rather than downflows 
as detected in our study. This difference may be due to the difference in phases of the flares as we detect the downflows in the late phase (4 minutes after the flare peak).


Using the gas density and pressure stratification obtained from NICOLE,  
under the assumption of hydrostatic equilibrium, we investigated the evolution of the total internal energy of the lower solar atmosphere 
covering the Ca {\sc{ii}} 8542 $\AA$ formation height.
Our analysis shows that the total energy of the chromosphere close to the flare maximum ($\sim$11:38 UT) 
is significantly increased in the range of optical depths 
log $\mathrm{\tau\sim-3.5-5.5}$ compared to the internal energy of the QS (Figure 6).  
A maximum enhancement was detected at $\mathrm{log~\tau\approx-5}$ 
where the total internal energy, integrated over a selected area of the flare, is a factor of 3 greater 
than the integral over the same area of a QS region. This internal energy changes reduces to the relaxed, QS state, after
approximately 15 minutes.
The total energy of the inverted box shown on the left panel of Figure~6
is estimated to be $\sim$10$^{24}$ erg close to the flare peak, and decreases linearly to the 
QS energy level $\sim\mathrm{6\times10^{23}}$ erg after 15 minutes (bottom right panel of Figure~6).
We note that hydrostatic equilibrium may not be a valid assumption for accurate estimations of the thermodynamical parameters, and hence the energy for a highly dynamical process such as a solar flare.  
However, we emphasise that modern inversion algorithms currently only use the assumption of 
hydrostatic equilibrium to obtain pressures and density stratifications and we believe that quantities 
such as the evolution of the energy ratio with time should be unaffected by this approximation.

To our knowledge, we have presented the first spectroscopic inversions, in NLTE, of the solar chromospheric line (Ca {\sc{ii}} 8542 {\AA}) 
to produce semi-empirical models of the flaring atmosphere. 
The temperature stratification obtained from the inversion are in good agreement with other semi-empirical NLTE flare 
models constructed without inversion codes, 
and with atmospheres obtained with forward modeling using radiative
radiative hydrodynamic simulations
\citep{1980ApJ...242..336M,2002A&A...387..678F,2005ApJ...630..573A,2015ApJ...813..125K,2016ApJ...827...38R}.
This suggests that NLTE inversions can be reliably applied to the flaring chromosphere. 


\begin{acknowledgements}

The research leading to these results has received funding from the European CommunityÕs Seventh Framework
Programme (FP7/2007Ð2013) under grant agreement No. 606862 (F-CHROMA). The Swedish 1-m Solar Telescope is operated on the island of La Palma 
by the Institute for Solar Physics (ISP) of Stockholm University at the Spanish Observatorio del Roque de los Muchachos of the Instituto de Astrof\'{\i}sica de Canarias.
The SST observations were taken within the Transnational Access and
Service Programme: High Resolution Solar Physics Network (EU-7FP
312495 SOLARNET). This research has made use of NASAÕs Astrophysics Data System.
This project made use of the Darwin Supercomputer of
the University of Cambridge High Performance Computing
Service (http:// www.hpc.cam.ac.uk/ ), provided by Dell
Inc. using Strategic Research Infrastructure Funding from
the Higher Education Funding Council for England and
funding from the Science and Technology Facilities Council.
The authors benefited from the excellent technical support of the SST staff, in particular
from the SST astronomer Pit S\"{u}tterlin.
This work was supported by the Science Grant Agency project VEGA
2/0004/16 (Slovakia) and by the Slovak Research and Development Agency
under the contract No. SK-AT-2015-0022.
This article was created by the realization of the project ITMS No. 26220120029,
based on the supporting operational Research and development
program financed from the European Regional Development Fund.
A.H. thank the Austrian FWF (project P 27765) for support. 
The work of T.V.Z. was supported by by the Austrian ÒFonds zur F\"orderung der wissenschaftlichen ForschungÓ (FWF) project P 28764-N27  
and by the Shota Rustaveli National Science Foundation project DI-2016-17.

\end{acknowledgements}

\bibliography{bibtex.bib}
\end{document}